\documentstyle[aps,manuscript]{revtex}
\begin{document}
\draft
\title{The Nystrom plus Correction Method
for Solving Bound State Equations in Momentum Space}
\author{Alfred Tang\thanks{atang@uwm.edu} 
and John W. Norbury\thanks{norbury@uwm.edu}}
\address{Physics Department, University of Wisconsin-Milwaukee,
P. O. Box 413, Milwaukee, WI 53201.}
\date{\today}
\maketitle

\begin{abstract}
A new method is presented for solving the momentum-space Schr\"{o}dinger
equation with a linear potential. 
The Lande-subtracted momentum space integral equation can be transformed
into a matrix equation by the Nystrom method.  The method produces 
only approximate eigenvalues in the cases of singular potentials such as
the linear potential.  The eigenvalues generated by the Nystrom method can be 
improved by calculating the numerical errors and adding the appropriate 
corrections.  The end results are more accurate eigenvalues than those 
generated by the basis function method.  The method is also shown to work
for a relativistic equation such as the Thompson equation.
\end{abstract}
\pacs{11.10.St, 11.10.Qr, 14.80.Dq}

\section{Introduction}
The momentum space Schr\"{o}dinger equation has a singular kernel for both
the Coulomb and linear potentials.  The Coulomb singularity is removed
with the Lande subtraction method~\cite{coul1,coul2}.  Previous 
work~\cite{maung93,kahana93,norbury91} showed how to remove the
singularity from the linear potential using a subtraction method with
basis functions.  A problem with this method is that one must guess a
suitable set of basis functions in advance.
In this paper, we show that Nystrom method~\cite{nr97} can solve the same 
problem more simply amd accurately.  We begin with
a review of the basis function method.  Then we introduce the Nystrom method
and apply it to the s-state momentum space Schr\"{o}dinger equation with a 
linear potential.  We use our new numerical results to show that the Nystrom 
plus correction method is more accurate than the basis function method.  At the
end, we generalize the
Nystrom method to higher angular momentum quantum numbers ($l>0$).

\section{The Basis Function Method}
We begin this paper with a discussion of the basis function method to give 
the proper theoretical motivation.
We shall use the simplest momentum space Schr\"{o}dinger equation 
to illustrate the principles of the numerical methods, which is the s-state
equation.  

The momentum space Schr\"{o}dinger equation is related to an 
integral equation of the Fredholm type
\begin{equation}
\int K(p,p') \phi(p') dp' = \lambda\phi(p).
\label{fredholm}
\end{equation}
Suppose that the wavefunction $\phi$ can be expanded in a set of basis
functions $\{g_{i}\}$, such that
\begin{equation}
\phi(p) = \sum^{N}_{i=1} C_{i}\,g_{i}(p),
\label{bf}
\end{equation}
where $C_{i}$ are constant coefficients.  Substitute Eq.~[\ref{bf}] 
into Eq.~[\ref{fredholm}] to obtain
\begin{equation}
\sum^{N}_{i=1}\int K(p,p')\,C_{i}g_{i}(p')\,dp'=
\lambda\sum^{N}_{i=1}C_{i}g_{i}(p).
\label{bi}
\end{equation}
Now multiply both sides of Eq.~[\ref{bi}] with $g_{j}(p)$ and integrate
over $p$ to symmetrize the equation over $i$ and $j$,
\begin{equation}
\sum^{N}_{i=1}C_{i} 
\underbrace{\int\int K(p,p')\,g_{i}(p')g_{j}(p)\,dp'dp}_{A_{ij}}
=\lambda\sum^{N}_{i=1}C_{i}
\underbrace{\int g_{i}(p)g_{j}(p)\,dp}_{B_{ij}},
\end{equation}
and the result is a matrix equation,
\begin{equation}
\sum^{N}_{i=1} A_{ij}C_{i} = \lambda \sum^{N}_{i=1} B_{ij}C_{i},
\end{equation}
where $C_{i}$ is the eigenvector and $\lambda$ the eigenvalue.
The indices $i$ and $j$ correspond to the quadrature points $p$ and $p'$.  
$N$ represents the number of mesh points.  
In the case of the momentum space
Schr\"{o}dinger equation with a Coulomb or linear potential, the kernel 
$\mathbf A$ is singular.  
A simple example is the momentum space Schr\"{o}dinger equation with a linear
potential in the $S$-state~\cite{maung93,norbury91},
\begin{equation}
\frac{p^{2}}{2\mu}\,\phi_{n0}(p)+\frac{\lambda_{L}}{\pi p^{2}}
\int^{\infty}_{0}\underbrace{\left[ {\eta^{2}\over p'p}Q''_{0}(y)+
Q'_{0}(y)\right]}_{V^{L}_{0}(p,p')}\,
\phi_{n0}(p')\, dp'
= E_{n0}\,\phi_{n0}(p),
\label{se1}
\end{equation}
where $y=(p^{2}+p'^{2})/2p'p$,
\begin{equation}
Q'_{0}(y)= p'p \left[ \frac{1}{(p+p')^{2} + \eta^{2}} -
\frac{1}{(p-p')^{2} + \eta^{2}} \right], \label{eq:Q'}
\end{equation}
and
\begin{equation}
\frac{\eta^{2}}{p'p}Q''_{0}(y)=\eta^{2} \left( p^{2}+p'^{2}+\eta^{2} \right)
\left[ \frac{1}{(p+p')^{2}+\eta^{2}}
-\frac{1}{(p-p')^{2}+\eta^{2}} \right]^{2}. \label{eq:Q''}
\end{equation}
Lande subtraction~\cite{maung93,norbury91,tang01} involves subtracting a 
zero term
\begin{equation}
\int^{\infty}_{0}\left[ {\eta^{2}\over p'p}Q''_{0}(y)+Q'_{0}(y)\right]\,dp'
=0
\end{equation}
from Eq.~[\ref{se1}] such that
\begin{equation}
\frac{p^{2}}{2\mu}\,\phi_{n0}(p)+\frac{\lambda_{L}}{\pi p^{2}}\int^{\infty}_{0}
\left[ {\eta^{2}\over p'p}Q''_{0}(y)+Q'_{0}(y)\right]
[\phi_{n0}(p')-\phi_{n0}(p)]\, dp' 
= E_{n0}\,\phi_{n0}(p).
\label{se2}
\end{equation}
Using Eqs.~[\ref{eq:Q'},\ref{eq:Q''}], the integral in Eq.~[\ref{se2}] for 
$p>0$ in the limit of $y\to 1$ can be shown to equal
\begin{equation}
\lim_{\eta\to 0}\,\lim_{p\to p'}\,
{\lambda_{L}\over\pi}\,\left[2\eta^{2}\left({1\over (p-p')^{2}+\eta^{2}}
\right)^{2}
-{1\over (p-p')^{2}+\eta^{2}}\right]\,(p-p')^{2}\,{d\phi_{n0}\over dp}=0.
\label{sing1}
\end{equation}
The order of the limits in Eq.~[\ref{sing1}] is important.
The reverse order will lead to the nonsensical result $\int Q'_{0}(y)\,dp'=0$. 
Next, in the limit of $p,p'\to 0$, $(p+p')^{2}=(p-p)^{2}$.  
By substituting 
this equality into Eqs.~[\ref{eq:Q'},\ref{eq:Q''}], it can be shown again that
the integral in Eq.~[\ref{se2}] vanishes for $p\to 0$ at $y=1$.  At the end, 
the integral vanishes at $y=1,\;\forall\, p$.  Away from the singularities, 
both integrands in the integral of Eq.~[\ref{se2}] are finite.  By taking 
$\eta\to 0$, the first integrand vanishes.  The final form of Eq.~[\ref{se2}] 
is
\begin{equation}
\frac{p^{2}}{2\mu}\,\phi_{n0}(p)+\frac{\lambda_{L}}{\pi p^{2}}\int^{\infty}_{0}
Q'_{0}(y)\,[\phi_{n0}(p')-\phi_{n0}(p)]\, dp' 
= E_{n0}\,\phi_{n0}(p),
\label{lin0}
\end{equation}
where $Q'_{0}(y)=1/(1-y^{2})$.
As mentioned before, $\phi$ is expanded in basis
functions, followed by integrating Eq.~[\ref{lin0}] over $p$ to generate a 
matrix equation.  The basis functions used in previous 
publications~\cite{maung93,norbury91} are
\begin{equation}
g^{A}_{i}(p)=exp\left[{-p^{2}i^{2}\over M}\right]
\label{bfa}
\end{equation}
and
\begin{equation}
g^{B}_{i}(p)={1\over (i/M)^{2}+p^{4}}
\end{equation}
respectively, where $M$ is the maximum number of basis functions used.  $M$
has a maximum because the code crashes when too many basis functions
are used.  The basis functions $g^{A}_{i}(p)$ and $g^{B}_{i}(p)$ have
$M=18$.

The singularity of the kernel is a major challenge in solving the integral
equation with a linear potential.  It was shown~\cite{maung93} that a simple 
pole remains even after
subtraction.  The role of the basis functions is to make possible the 
evaluation of the Cauchy principal value of the subtracted integral using the
Sloan method~\cite{sloan}.  To illustrate the Sloan method, we suppose that 
$f(x)$ has a simple pole such that
\begin{equation}
f(x)={g(x)\over x}
\end{equation}
where $g(x)$ is regular.  The Cauchy principal value of the subtracted 
integral of $f(x)$ can be evaluated if the range of integration is symmetric.
For example, the numerical integration of 
\begin{equation}
\int^{1}_{-1} {g(x)-g(0)\over x}\,dx
\end{equation}
yields the Cauchy principal value because the point $x=0$ is skipped when 
quadrature points are generated in the symmetric interval $(-1,1)$.  The
subtraction term has zero contribution since
\begin{equation}
\int^{1}_{-1} {1\over x}\,dx = 0.
\end{equation}
The purpose of this term is to justify the existence of the Cauchy principle
value and to reduce numerical errors.  In order to apply the Sloan method to
Eq.~[\ref{lin0}], the integration variable is transformed from $p$ to
$x$ such that $x$ is centered at zero and its range is symmetric.

In the case of the
Coulomb potential, the kernel has a logarithmic singularity,
\begin{equation}
Q_{0}(y)={1\over 2}\ln\left| {y+1\over y-1}\right|,
\end{equation}
which is completely removed by Lande subtraction~\cite{coul1,maung93} because 
no simple pole remains after the subtraction.

The key to the success of the basis function method is the availability of
a suitable set of basis functions for a particular problem.  Unfortunately
there is no \emph{a priori} reason why the same set of basis functions will 
work in every situation.  For this reason, it may be advantageous to have a 
method (such as the Nystrom method) that does not depend on a choice of the 
basis functions.

\section{The Nystrom Method}
In general, an integral equation of the Fredholm type
\begin{equation}
G(p)\phi(p)+
\int^{\infty}_{0} F(p,p')\, \phi(p') \,dp' = \lambda\,\phi(p) \label{inteq}
\end{equation}
can be rewritten as a matrix equation as
\begin{equation}
\sum^{N}_{j=1} K_{ij}\,\phi_{j} \equiv
\sum^{N}_{j=1} \left( G_{ii}+F_{ij}\right) \,\phi_{j} = \lambda\,\phi_{i},
\end{equation}
where $K_{ij}$ is the kernel and $i$ and $j$ are now indices corresponding
to $p$ and $p'$.
Instead of integrating over $p$ from 0 to $\infty$, we integrate
over $x$ from $-1$ to 1.  Transform $x_{i}$ to $p_{i}$ by the 
transformation
\begin{equation}\label{trans}
p(x)=\tan\left( {1+x\over 4}\,\pi \right).
\end{equation}
The mesh points $x_{i}$ and the weights $wt_{i}$ are generated by the
gaussian quadrature rule using the routine \texttt{gauleg} from {\em Numerical
Recipes}~\cite{nr97}.
In order to integrate along $x\in[-1,1]$ instead of 
$p\in[0,\infty)$, Eq.~[\ref{inteq}] is transformed as
\begin{equation}
G(x)\phi(x)+
\int^{1}_{-1} F(x,x') \,\phi(x') \,{dp'\over dx'}\,dx' = \lambda\,\phi(x)
\end{equation}
Changing the dummy variable inside the integral and substituting the
differentiation of Eq.~[\ref{trans}] with
\begin{equation}
dp = {\pi\over 4}\,\sec^{2}\left( {1+x\over 4}\,\pi \right)\, dx
={\pi\over 4}\,\left(1+p^{2} \right)\, dx
\end{equation}
gives
\begin{equation}
\frac{p_{i}^{2}}{2\mu}\,\phi_{i}+\frac{\lambda_{L}}{4 p_{i}^{2}}
\int^{1}_{-1} Q'_{0}(y\ne 1) [ \phi_{j}-\phi_{i} ]\,
\,\sec^{2}\left( {1+x_{j}\over 4}\,\pi \right)\,dx_{j} 
= E\,\phi_{i}.
\label{pre_matrix}
\end{equation}
Eq.~[\ref{pre_matrix}] can now be written as a matrix equation,
\begin{eqnarray}
&&\frac{p_{i}^{2}}{2\mu}\,\phi_{i}+\frac{\lambda_{L}}{4 p_{i}^{2}}
\sum^{N}_{j=1}Q'_{0}(y\ne 1)\phi_{j}\,
\sec^{2}\left( {1+x_{j}\over 4}\,\pi \right)\,wt_{j} \nonumber\\
&&\quad -\frac{\lambda_{L}}{4 p_{i}^{2}}\,
\phi_{i}\sum^{N}_{k=1}Q'_{0}(y\ne 1)\,
\sec^{2}\left( {1+x_{k}\over 4}\,\pi \right)\,wt_{k}
= E\,\phi_{i}.
\label{matrix}
\end{eqnarray}
The left hand side of Eq.~[\ref{matrix}] is the kernel times the eigenvector
and the right hand side is the eigenvalue times the eigenvector.  The sum 
over $k$ is independent of the eigenvector, which is just a scalar.
The terms on the left that have only one index $i$ belong to
the diagonal elements $K_{ii}$.
The terms with mixed indices $i$ and $j$ make up the off-diagonal elements,
$K_{ij}$.  More explicitly, the matrix elements of the kernel are
\begin{eqnarray}
K_{ii} &=& \frac{p_{i}^{2}}{2\mu} - \frac{\lambda_{L}}{4 p_{i}^{2}}
\sum_{k}Q'_{0}(y \ne 1)\,\left(1+p^{2}_{k}\right)
\,wt_{k}, \label{kii}\\
K_{ij} &=& \frac{\lambda_{L}}{4 p_{i}^{2}}\,Q'_{0}(y)\,
\left(1+p^{2}_{j}\right)\,wt_{j}, \quad (i\ne j).\label{kij}
\end{eqnarray}

So far the kernel $\mathbf K$ is asymmetric under the interchange of $i$
and $j$.  We can improve the 
stability and the efficiency of the numerical solutions by symmetrizing 
Eq.~[\ref{matrix}].  We do so by multiplying the equation 
with $p_{i}^{2}\,\left(1+p^{2}_{i}\right)$.  It will 
change the original matrix equation
\begin{equation}
{\mathbf K}\cdot{\mathbf x}=\lambda\,{\mathbf x}
\end{equation}
to an equivalent matrix equation
\begin{equation}
{\mathbf K'}\cdot{\mathbf x}=\lambda\,{\mathbf C}\cdot{\mathbf x}, \label{meq0}
\end{equation}
where $\mathbf C$ is a diagonal matrix and ${\mathbf K'}={\mathbf C}\cdot
{\mathbf K}$.
If ${\bf C}$ is positive definite, meaning
\begin{equation}
{\bf x}^{T}\cdot {\bf C}\cdot {\bf x} \geq 0, \quad
\forall\; {\rm vectors}\;{\bf x},
\end{equation}
then ${\bf C}$ can be Cholesky-decomposed as
\begin{equation}
{\bf C} = {\bf L}\cdot {\bf L}^{T},
\end{equation}
where ${\bf L}$ is a unique lower triangular matrix.  
The reason for performing Cholesky decomposition is that the new matrix
\begin{equation}
{\bf K''} \equiv {\bf L}^{-1}\cdot {\bf K'}\cdot ({\bf L}^{-1})^{T}
\end{equation}
is real symmetric and yields the same eigenvalues as Eq.~[\ref{meq0}].
In the case of $C_{ii}=p^{2}_{i}\,\left(1+p^{2}_{i}\right)$, 
$\mathbf C$ is guaranteed 
to be positive definite.  After symmetrization, Eq.~[\ref{kii}] does not 
change ($K''_{ii}=K_{ii}$) while Eq.~[\ref{kij}] becomes
\begin{eqnarray}
K''_{ij} &=& \frac{\lambda_{L}}{4p_{i}p_{j}}\,Q'_{0}(y)\,
\sqrt{\left(1+p^{2}_{i}\right)\left(1+p^{2}_{j}\right)wt_{i}wt_{j}}, 
\quad (i\ne j).
\end{eqnarray}

The eigenvalues of $\mathbf K''$ can be calculated by using standard packages
such as EISPAK.  In this paper, we use the \texttt{tred2} and \texttt{tqli} 
routines in \emph{Numerical Recipes}~\cite{nr97}.

\section{The Correction Method}
Maung, Norbury and Kahana~\cite{maung93,norbury91} have shown that the
subtraction method
does not completely remove the singularity at $y=1$.  There is a residual
simple pole term
\begin{equation}
-{4\lambda_{L}\over \pi} \, {d\phi_{n0}\over dp}\int^{\infty}_{0}
{p'^{2}\over(p'+p)^{2}(p'-p)}\,dp'
\label{error}
\end{equation}
remaining after the subtraction.  The basis function method evaluates the 
Cauchy principal value by the Sloan method as described in Section 2.
The Sloan method eliminates
the simple pole term by integrating symmetrically around the singularity.
Symmetrical integration involves splitting the range of integration is into 
two intervals,
\begin{equation}
\int^{\infty}_{0}dp'=\int^{2p}_{0}dp'+\int^{\infty}_{2p}dp'.
\label{sym}
\end{equation}
The singularity at $p=p'$ is contained in the first term on the right hand
side of Eq.~[\ref{sym}] which is assigned a symmetric transformation rule 
$(dp'/dx)_{1}$.  The second term generally has a different transformation
rule $(dp'/dx)_{2}$ because it is mapping between two different sets, namely 
$(2p,\infty)\to(1,M]$ (for some real number $M$), such that
\begin{equation}
\int^{\infty}_{0}dp'\to\int^{1}_{-1}\left({dp'\over dx'}\right)_{1}\,dx'+
\int^{M}_{1}\left({dp'\over dx'}\right)_{2}\,dx'.
\end{equation}
Notice that the division of the range of integration moves with $p$.  If two
transformation rules are used with a moving division,
each row (column) of the kernel has a different way to map $[0,\infty)$
to $[-1,M]$.  But the eigenvector $\phi(p)$ must be mapped to $\phi_{i}$ in a 
unique way.  This mismatch between the mappings of the kernel and the 
eigenvector does not affect the basis function method (see Eq.~[2.24] of
reference~\cite{maung93})
\begin{eqnarray}
&&\sum^{N}_{i=1}C_{i}\left[\int^{\infty}_{0}{p^{4}\over 2\mu}\,g_{j}(p)
g_{i}\,dp
+{\lambda_{L}\over\pi}\int^{\infty}_{0}\int^{\infty}_{0}Q'_{0}(y)g_{j}(p)
\left[g_{i}(p')-g_{i}(p)\right]\,dp'dp\right]\nonumber\\
&=&E\sum^{N}_{i=1}C_{i}\int^{\infty}_{0}p^{2}g_{j}(p)g_{i}(p)\,dp
\end{eqnarray}
because the 
eigenvector $C_{i}$ is an $N$-tuple of the coefficients of the basis function 
expansion of the
wavefunction $\phi(p)$ and is independent of the transformation rules.
In the case of the Nystrom method, the problem is real, at least for 
the range of integration that we are interested in.  Therefore
we cannot evaluate the Cauchy principal value by symmetric integration in the 
Nystrom method.  In other words, a new method is needed to treat the 
errors arising from the simple pole term.

So far the error term Eq.~[\ref{error}] is not included in the Nystrom 
kernel in our derivation and is contributing to the
errors of the eigenvalues.  Since the error term
Eq.~[\ref{error}] involves $d\phi/dp$, we associate it with the error of the 
wavefunction
\begin{equation}
\Delta\phi=\Delta p{d\phi\over dp}\sim{1\over N}\,{d\phi\over dp},
\end{equation}
where the mesh size $\Delta p$ has an $N^{-1}$ dependence.
This fact leads to an estimate of the $N$ dependence of the error of the 
eigenvalue, $\Delta E$.  Let the approximate eigenvector to be $\phi'$
and the approximate eigenvalue $E'$.  It is reasonable to say
that an approximate kernel $K$ acting on an approximate eigenvector $\phi'$
yields an approximate eigenvalue $E'$ as in
\begin{eqnarray}
&&K\phi'=E'\phi'\\
&\Rightarrow& K(\phi+\Delta\phi)=(E+\Delta E)(\phi+\Delta\phi).
\end{eqnarray}
It is easy to see that
\begin{eqnarray}
\Delta E&\simeq&(K-E){\Delta\phi\over\phi}\nonumber\\
&=&\left({K-E\over\phi}\right)\,{d\phi\over dp}\,\Delta p\nonumber\\
&=&\epsilon\,{1\over N}
\end{eqnarray}
It is safe to assume that $(K-E)\,d\phi/dp<<1$.  $\phi^{-1}$ can be 
interpreted as the normalization.  The product of all of the
pseudo-constants is labelled as the coefficient $\epsilon$.  The approximate 
eigenvalue $E'$ produced in the background of Eq.~[\ref{error}] is related to
the true eigenvalue $E$ by
\begin{equation}
E_{n}'=E_{n}+\epsilon\,f_{n}(N),
\end{equation}
where $n$ is the principal quantum number, $\epsilon$ a constant and 
$f_{n}(N)$ is a function approximately equals to $N^{-1}$.  In 
general, $f_{n}(N)$ varies slightly depending on the type of integral equation 
and the potential.  As a first order approximation, assume that
\begin{equation}
f_{n}(N)=N^{-1-\alpha\,(n-1)}.
\label{fn}
\end{equation}
The exponent of Eq.~[\ref{fn}] is a first order Taylor series expansion of
some negative unity function around $n=1$.  The constant $\alpha$ is always 
taken to be small.  More particularly, choose an $\alpha$ such that the 
variance of $E_{n}$ and $\epsilon$ and $\chi^{2}$ are minimized in the linear
fit.  Finally the refinement of an eigenvalue involves generating a set of 
$E_{n}'$ for various $N$ by the Nystrom method and then extrapolating
$E_{n}$ by a $\chi^{2}$ linear fit in the graph of
$E_{n}$ versus $f_{n}(N)$.  In the case of Eq.~[\ref{lin0}],
$\alpha=0.004$ is an optimal choice.  The numerical results are explained in 
Section 6.

The order of the Nystrom algorithm is derived from those of \texttt{tred2} and 
\texttt{tqli}, which is ${\mathcal O}(N^{2})$~\cite{nr97}, compared with the 
basis function's ${\mathcal O}(M^{2}N)$, which comes from the 
product of the size of the matrix $M^{2}$ and the number of integration
mesh points.  $N$ is typically around 1000 and $M$ is 20.  The basis function
method is generally more efficient than the Nystrom method.  However, for 
any given set of
basis functions, the accuracy of the eigenvalues cannot be improved arbitrarily
by increasing the number of basis functions because $M$ is bounded from above
due to numerical errors.
The prospect of improving the accuracy of the basis function algorithm depends
on the availability of a set of more suitable basis functions for a
specific problem.  In the
case of the Nystrom plus correction method, accuracy is optimized 
automatically by the correction scheme.  The numerical results 
obtained by the Nystrom and basis function methods are quoted with optimal
accuracy in this paper.

\section{Exact S-state Solution}
The eigenvalue of Eq.~[\ref{lin0}] can be solved exactly in configuation space.
We shall use the analytic results to check our numerical results.  The
non-relativistic Schr\"{o}dinger equation can be written as
\begin{equation}
\left( {d^{2}\over dr^{2}} + {2\over r}\,{d\over dr} \right)\,R -
2\mu[\lambda_{L}\,r - E]\,R = 0. \label{linear_r}
\end{equation}
Let $S\equiv r\,R$, then Eq.~[\ref{linear_r}] can be simplified as
\begin{equation}
{d^{2}\over dr^{2}}S - 2\mu[\lambda_{L}\,r - E]S = 0.
\end{equation}
Define a new variable
\begin{equation}
x\equiv \left( {2\mu\over \lambda_{L}^{2}} \right)^{1\over 3}
[\lambda_{L}r - E], \label{linear_s}
\end{equation}
such that Eq.~[\ref{linear_r}] can be transformed as
\begin{equation}
S'' - xS = 0,
\end{equation}
which is the Airy equation.  The solution which satisfies the boundary
condition $S\to 0$ as $x\to \infty$ is the Airy function
${\rm Ai}(x)$.  It is easy to show that the eigen-energy formula is
\begin{equation}
E_{n} = -x_{n}\,\left( {\lambda_{L}^{2}\over 2\mu} \right)^{1\over 3},
\end{equation}
where $x_{n}$ is the $n$-th zero of the Airy function counting from
$x=0$ along $-x$.  
In reference~\cite{norbury91}, the values
$\lambda_{L} = 5$ and $\mu = 0.75$ are used.  In this case, the eigen-energy 
formula is
\begin{equation}
E_{n} = -2.554364772\,x_{n}.
\end{equation}

\section{Numerical Results for the S-state}
The accuracy of the Nystrom plus correction method is sensitive to
the range of $N$.  In this paper, increments of 100 in the range of
$100\le N \le 1400$ are used.  The reason for this choice is that there are 
not enough spacings between the eigenvalues for $N<100$ and 
for $N>1500$ the numerical noise begins to corrupt the monotonic convergent
behavior of the eigenvalues.  The
correct eigenvalues are extrapolated from these numerical data by a
$\chi^{2}$ linear fit as described in Section 4.
The exact $S$-state eigenvalues are tabulated against the numeical results 
obtained by the basis function method and the Nystrom plus corrections method 
in Table~[\ref{compare_tab_1}].  It shows that the numerical results obtained
by the Nystrom method plus corrections are more accurate than the
results obtained by the basis function method.
  
The kernel written for the
non-relativistic Schr\"{o}dinger equation can be easily generalized to that
of the relativistic 2-body Thompson equation in the center-of-mass
frame by the replacement
\begin{equation}
{p^{2}\over 2\mu}\to 2\left(\sqrt{p^{2}+m^{2}}-m\right),
\end{equation}
where $\mu$ is the reduced mass and $m$ is the mass of each of the 
two equal mass elementary particles.
The numerical results obtained using the Thompson equation is compared
against those using the non-relativistic Schr\"{o}dinger equation in
Table~[\ref{tnr}] calculating to 2 decimal places.  Our new results are exactly
the same as the previous results obtained in reference~\cite{maung93} that
uses basis functions $g^{A}_{i}(p)$ from Eq.~[\ref{bfa}].

\section{$l\ne 0$ Kernels}
The $l\ne 0$ kernels for the linear and Coulomb potentials contain
the Legendre function of the second kind $Q_{l}(y)$ and its derivative
respectively.  There are several mathematical issues
that need to be addressed before constructing the $l\ne 0$ kernels.
First of all, the definition of
\begin{equation}
y\equiv {p^{2}+p'^{2}\over 2p'p}={1\over 2}\,\left({p\over p'}+{p'\over p}
\right)
\end{equation}
is easily seen to yield $y>1$ for $p,p'>0$.
In reference~\cite{maung93}, Maung \emph{et al.} use the Legendre identity
\begin{eqnarray}
Q_{l}(y) &=& P_{l}(y)Q_{0}(y)-w_{l-1}(y), \nonumber\\
w_{l-1}(y) &=& \sum^{l}_{m=1}{1 \over m}\,P_{l-m}(y)P_{m-1}(y),
\label{ql}
\end{eqnarray}
which is valid for $-1\le y\le 1$~\cite{copson,stegun} but
can be extended to $y>1$ by analytic continuation~\cite{crc}.  $Q'_{l}(y)$
is easily obtained by straightforward differentiation.
The derivative of Legendre polynomial can be calculated from
one of the recurrence formulas,
\begin{equation}
{dP_{l}(y)\over dy}=y\,{dP_{l-1}(y)\over dy}+lP_{l-1}(y),
\end{equation}
which can be computed numerically by a recursive call.  The Legendre
function can be generated by modifying the routine \texttt{plgndr} in
{\em Numerical Recipes}~\cite{nr97} to allow $y>1$.  The accuracy of 
Eq.~[\ref{ql}] and its derivatives are generally sufficient. Slightly more 
accurate results can be obtained by the explicit evaluation of the Neumann 
integral,
\begin{equation}
Q_{l}(y)={1 \over 2}\int^{1}_{-1}{1\over (y-t)}\,P_{l}(t)\,dt,
\label{neumann}
\end{equation}
with derivative
\begin{equation}
Q'_{l}(y)=-{1 \over 2}\int^{1}_{-1}{1\over (y-t)^{2}}\,P_{l}(t)\,dt.
\label{neumannder}
\end{equation}
The first few $Q_{l}(y)$ are
\begin{eqnarray}
Q_{0}(y)&=&{1\over 2}\,\ln{y+1\over y-1},\label{q_0}\\
Q_{1}(y)&=&{1\over 2}\,y\ln{y+1\over y-1}-1,\\
Q_{2}(y)&=&{1\over 4}\left( 3y^{2}-1\right)\,\ln{y+1\over y-1}-{3\over 2}\,y,\\
Q_{3}(y)&=&{1\over 4}\left( 5y^{3}-3y\right)\,\ln{y+1\over y-1}-
{5\over 2}\,y^{2}+{2\over 3},\\
Q_{4}(y)&=&{1\over 16}\left( 35y^{4}-30y^{2}+3\right)\,\ln{y+1\over y-1}
-{35\over 8}\,y^{3}+{55\over 24}\,y,\\
Q_{5}(y)&=&{1\over 16}\left( 63y^{5}-70y^{3}+15y\right)\,\ln{y+1\over y-1}
-{63\over 8}\,y^{4}+{49\over 8}\,y^{2}-{8\over 15}.
\end{eqnarray}
$Q'_{l}(y)$ can be obtained by the direct differentiation of $Q_{l}(y)$,
such that
\begin{eqnarray}
Q'_{0}(y)&=&{1\over 1-y^{2}},\\
Q'_{1}(y)&=&{y\over 1-y^{2}}-{1\over 2}\,\ln{y-1\over y+1},\\
Q'_{2}(y)&=&{1\over 1-y^{2}}-{3\over 2}\,y\ln{y-1\over y+1}-3,\\
Q'_{3}(y)&=&{y\over 1-y^{2}}-{15y^{2}-3\over 4}\,\ln{y-1\over y+1}
-{15\over 2}\,y,\\
Q'_{4}(y)&=&{1\over 1-y^{2}}-{35y^{3}-15y\over 4}\,
\ln{y-1\over y+1}-{35\over 2}\,y^{2}+{5\over 3},\\
Q'_{5}(y)&=&{y\over 1-y^{2}}-{315y^{4}-210y^{2}+15\over 16}\,\ln{y-1\over y+1}
-{315\over 8}\,y^{3}+{105\over 8}\,y.\label{qp_5}
\end{eqnarray}
As $y\to\infty$, it is easily seen that $Q_{0}(y)=Q'_{0}(y)\to 0$.  This
limit is true for all $Q_{l}(y)$ and $Q'_{l}(y)$ from applying the 
L'Hopital rule.  Unfortunately
straightforward numerical calculation of $Q_{l}(y)$ and $Q'_{l}(y)$ by using
Eqs.~[\ref{q_0}-\ref{qp_5}] leads to serious numerical errors as
$y\to\infty$.  At the same time, it is observed that the numerical integration 
of Eq.~[\ref{neumann},\ref{neumannder}] are reasonably accurate in the same
regime.  Therefore the two representations are combined to minimize 
numerical error by using the Neumann integrals for $y>y_{0}$ and the 
explicit formulas for $y\le y_{0}$.  Our codes use $y_{0}=50$.

The subtracted momentum space NRSE with a linear potential is given in 
reference~\cite{maung93}, which can be simplified as
\begin{eqnarray}
&& {p^{2}\over 2\mu}\phi_{nl}(p) + {\lambda_{L}\over \pi p^{2}}
\int^{\infty}_{0} Q'_{l}(y)\,\phi_{nl}(p')\,dp' - 
{\lambda_{L}\over \pi p^{2}}\,\phi_{nl}(p)
\int^{\infty}_{0} Q'_{0}(y)\,dp'\nonumber\\
&& \quad - {\lambda_{L}\over \pi p}\, {l(l+1)\over 2}\,\phi_{nl}(p)
\int^{\infty}_{0} {Q_{0}\over p'}\,dp'
+ {\lambda_{L}\,\pi\over p}\,{l(l+1)\over 4}\,\phi_{nl}(p)
= E_{nl}\phi_{nl}(p).
\end{eqnarray}
The matrix elements of a symmetric kernel for arbitrary $l$ are
\begin{eqnarray}
K_{ii} &=& \frac{p_{i}^{2}}{2\mu} - \frac{\lambda_{L}}{4 p_{i}^{2}}
\sum_{k}Q'_{0}(y \ne 1)\,\left(1+p^{2}_{k}\right)\,wt_{k}, \nonumber\\
&&\quad - \frac{\lambda_{L}}{4 p_{i}}\,{l(l+1)\over 2}
\sum_{k}{Q_{0}(y \ne 1)\over p_{k}}\,\left(1+p^{2}_{k}\right)\,wt_{k}
+\frac{\lambda_{L}\,\pi}{4 p_{i}}\,l(l+1)\nonumber\\
&&\quad - \frac{\lambda_{L}}{4 p_{i}^{2}}\,w'_{l}(1)
\,\left(1+p^{2}_{i}\right)\,wt_{i}\\
K_{ij} &=& \frac{\lambda_{L}}{4 p_{i}p_{j}}\,Q'_{l}(y)\,
\sqrt{\left(1+p^{2}_{i}\right)\left(1+p^{2}_{j}\right)\,wt_{i}\,wt_{j}}, 
\quad (i\ne j).
\end{eqnarray}
Despite our method to control numerical noise, numerical errors still
manifest themselves in the form of spurious large negative eigenvalues for
$l\ge 8$.  Fortunately the rest of the positive eigenvalues are accurate.
Some sample eigenvalues for $0\le l\le 5$ are shown in Table~[\ref{linear}]
which also compares the eigenvalues generated by both the $p$-space and
$r$-space codes.  The $r$-space eigenvalues are calculated by solving 
NRSE using the relaxation method~\cite{tns}.

The Lande-subtraced momentum space NRSE equation with a Coulomb potential
is also given in reference~\cite{maung93} and is simplified as
\begin{eqnarray}
&&{p^{2}\over 2\mu}\phi_{nl}(p) + {\lambda_{C}\over \pi p}
\int^{\infty}_{0} Q_{l}(y)\,\phi_{nl}(p')\,p'\,dp'\nonumber\\
&&\quad -{\lambda_{C}\over \pi}\,p\,\phi_{nl}(p)
\int^{\infty}_{0} {Q_{0}(y)\over p'}\,dp'
+ {\lambda_{C}\,\pi\over 2}\,p\,\phi_{nl}(p)
= E_{nl}\phi_{nl}(p).
\end{eqnarray}
The kernel of a Coulomb potential can be symmetrized  the same way as that 
of a linear potential.  The matrix elements are
\begin{eqnarray}
K_{ii} &=& \frac{p_{i}^{2}}{2\mu} -\frac{\lambda_{C}}{4}\,p_{i}\,
\sum_{k}{Q_{0}(y \ne 1)\over p_{k}}\,\left(1+p^{2}_{k}\right)\,wt_{k}
+{\lambda_{C}\,\pi\,p_{i}\over 2}\nonumber\\
&&\quad - \frac{\lambda_{C}}{4}\,w_{l}(1)
\,\left(1+p^{2}_{i}\right)\,wt_{i}\\
K_{ij} &=& \frac{\lambda_{C}}{4}\,Q_{l}(y)\,
\sqrt{\left(1+p^{2}_{i}\right)\left(1+p^{2}_{j}\right)\,wt_{i}\,wt_{j}}, 
\quad (i\ne j).
\end{eqnarray}
The correction method which we have developed for the linear potential cannot
be used in the Coulomb case.  The only available technique
of refining the eigenvalues of a Coulomb potential is by the way of increasing 
the number of mesh steps $N$.  Some sample eigenvalues are shown in 
Table~[\ref{coulomb}].  Since both the linear and Coulomb potentials can be
symmetrized using the same formalism, we can easily splice the two kernels
together to calculate the eigenvalues of the Cornell (linear plus
Coulomb) potential
\begin{equation}
V(r)={\lambda_{C}\over r}+\lambda_{L}\,r.
\end{equation}
It is not surprising that the correction method derived for the linear
potential may also work for the Cornell potential because we expect that
the error of the Cornell potential is dominated by the error of the
linear potential term.  But it is a surprise that the correction
method works more accurately with the Cornell potential than the
linear potential as evidenced by vanishingly small variance and $\chi^{2}$.  

\section{Conclusion}
The basis function method requires {\em a priori} knowledge of the 
eigenfunctions in order to pick out an appropriate set of basis 
functions.  The advantage of the Nystrom method is that no such prior knowledge
of the eigenfunctions is needed.  The kernel constructed by the Nystrom 
method is also much simpler than that by the basis function method.
The eigenfunctions can be generated by the same Nystrom routines that compute 
the eigenvalues.  The Nystrom plus correction is more accurate than the basis 
function method in the cases studied in this paper.  In other words, the new
method has all of the advantages--elegance, accuracy and 
versatility.  In addition, the kernel of the relativisitc and
non-relativistic equation of motion with the Coulomb and linear potential can 
be symmetrized in exactly the same manner.  It allows the calculation of the
eigenvalues of a Cornell potential readily.  Since the 
Nystrom method can be generalized for higher $l$'s, we can use it to calculate 
the Regge trajectories.  Since the Thompson
equation which we have solved is a 2-body equation, we can use it to analyze
the experimental meson Regge trajectories~\cite{tang}.  This will be
pursued in later work.

\acknowledgments
We thank Prof. Khin Maung Maung for his helpful comments and 
George Nill and Daniel R. Shillinglaw for their participation.

\begin{table}
\caption{Comparisons of eigen-energies in GeV of the non-relativistic
Schr\"{o}dinger equation with a linear potential between the Nystrom
method and the basis Function (BF) method.  The basis functions being
referred to here are $g^{B}_{i}(p)=[(i/M)^{2}+p^{4}]^{-1}$.
The values of $l=0$, $\lambda_{L}=5 \rm\, GeV$ and $\mu = 0.75 \rm\, GeV$ 
are used.}
\begin{tabular}{crrrrrr}
 && Nystrom &&& BF & Exact \\
$n$ & $N=100$ & $N=700$ & $N=1400$ & Corrected &&\\
\tableline
1  & 5.899211 & 5.961921 &  5.967339 & 5.972379 & 5.972 & 5.972379 \\
2  & 10.268443 & 10.417386 & 10.430047 & 10.442010 & 10.443 & 10.442114 \\
3  & 13.767781 & 14.054263 & 14.078517 & 14.101276 & 14.104 & 14.101524 \\
4  & 16.784747 & 17.258395 & 17.297500 & 17.335360 & 17.335 & 17.335728 \\
5  & 19.467512 & 20.177458 & 20.234722 & 20.291708 & 20.293 & 20.292215 \\
6  & 21.891635 & 22.887999 & 22.967933 & 23.046820 & 23.053 & 23.047142 \\
7  & 24.101339 & 25.435892 & 25.541743 & 25.646532 & 25.648 & 25.646268 \\
8  & 26.124257 & 27.851711 & 27.986463 & 28.121481 & 27.947 & 28.120787 \\
9  & 27.977844 & 30.156480 & 30.323418 & 30.493311 & 30.194 & 30.488938 \\
10 & 29.672260 & 32.366010 & 32.568895 & 32.778297 & 33.340 & 32.769375 \\
\end{tabular}
\label{compare_tab_1}
\end{table}

\newpage

\begin{table}
\caption{Comparisons of the ratios of Eigen-energies $E_{n+1}/E_{1}$ using
the Thompson equation (TE) and the non-relativistic Schr\"{o}dinger equation
(NRSE) using $l=0$ and $\lambda_{L}=0.2\,\rm GeV^{2}$.  Mass is measured in
GeV.}
\begin{tabular}{crrr}
$n$ & TE & NRSE & Mass \\
\tableline
1  & 1.72 & 1.75 & 1.5 \\
2  & 2.30 & 2.36 & 1.5 \\
3  & 2.80 & 2.90 & 1.5 \\
\tableline
1  & 1.67 & 1.75 & 0.5 \\
2  & 2.18 & 2.36 & 0.5 \\
3  & 2.62 & 2.90 & 0.5 \\
\tableline
1  & 1.63 & 1.75 & 0.3 \\
2  & 2.11 & 2.36 & 0.3 \\
3  & 2.51 & 2.90 & 0.3 \\
\end{tabular}
\label{tnr}
\end{table}

\newpage

\begin{table}
\caption{Eigen-energies in GeV of the non-relativistic
Schr\"{o}dinger equation in momentum space (pNRSE) compared with those
in the configuration space (rNRSE) and the relativistic 
Thompson equation in momentum space (TE).  The
$r$-space Thompson equation result is not available.
The values of $n=1$, $\lambda_{L}=5 \rm\, GeV$ and $\mu = 0.75 \rm\, GeV$ 
are used.}
\begin{tabular}{crrrrr}
&& pNRSE &&& rNRSE\\
$l$ & $N=100$ & $N=700$ & $N=1400$ & Corrected & Approx.\\
\tableline
0 & 5.899211 & 5.961921 & 5.967339 & 5.972379 & 5.9719 \\
1 & 8.528725 & 8.577713 & 8.582413 & 8.586002 & 8.5850 \\
2 & 10.823099 & 10.847533 & 10.849675 & 10.851526 & 10.8514 \\
3 & 12.917124 & 12.904221 & 12.902815 & 12.902117 & 12.9020 \\
4 & 14.874248 & 14.812422 & 14.805462 & 14.801358 & 14.9790 \\
5 & 16.730585 & 16.606651 & 16.597636 & 16.586361 & 16.5845 \\
\tableline
&& TE &&&\\
$l$ & $N=100$ & $N=700$ & $N=1400$ & Corrected & NA \\
\tableline
0 & 5.859885 & 5.914287 & 5.919054 & 5.923117 &\\
1 & 8.164379 & 8.202282 & 8.205185 & 8.208610 &\\
2 & 10.053574 & 10.067261 & 10.068464 & 10.069762 &\\
3 & 11.700322 & 11.680163 & 11.678063 & 11.676817 &\\
4 & 13.185124 & 13.121767 & 13.116634 & 13.111239 &\\
5 & 14.553134 & 14.437612 & 14.427702 & 14.418173 &\\
\end{tabular}
\label{linear}
\end{table}

\newpage

\begin{table}
\caption{Eigen-energies in eV of the hydrogen atom according to the 
non-relativistic Schr\"{o}dinger equation with $n=1$.}
\begin{tabular}{crrrr}
$l$ & $N=100$ & $N=1400$ & $N=3000$ & Exact \\
\tableline
0 & -25.286631 & -13.600349 & -13.598508 & -13.598289 \\
1 & -4.579043 & -3.400415 & -3.399659 & -3.399572 \\
2 & -1.463504 & -1.511499 & -1.510980 & -1.510921 \\
3 & -0.634523 & -0.850358 & -0.849940 & -0.849893 \\
4 & -0.329730 & -0.544332 & -0.543972 & -0.543932 \\
\end{tabular}
\label{coulomb}
\end{table}

\end{document}